\documentclass[aip, apl, reprint, amsmath,amssymb]{revtex4-1}

\usepackage{graphicx}
\usepackage{dcolumn}
\usepackage{bm}

\usepackage{epstopdf} 

\begin{document}
\title{Measurement and simulation of anisotropic magnetoresistance in single GaAs/MnAs core/shell nanowires}
\author{J. Liang}
\author{J. Wang}
\affiliation{Materials Research Institute, The Pennsylvania State University, University Park, Pennsylvania 16802, USA}\affiliation{Department of Physics, The Pennsylvania State University, University Park, Pennsylvania 16802, USA}
\author{A. Paul}
\affiliation{Materials Research Institute, The Pennsylvania State University, University Park, Pennsylvania 16802, USA}\affiliation{Department of Materials Science and Engineering, The Pennsylvania State University, University Park, Pennsylvania 16802, USA}
\author{B.J. Cooley}
\author{D. W. Rench}
\affiliation{Materials Research Institute, The Pennsylvania State University, University Park, Pennsylvania 16802, USA}\affiliation{Department of Physics, The Pennsylvania State University, University Park, Pennsylvania 16802, USA}
\author{N. S. Dellas}
\author{S. E. Mohney}
\author{R. Engel-Herbert}
\affiliation{Materials Research Institute, The Pennsylvania State University, University Park, Pennsylvania 16802, USA}\affiliation{Department of Materials Science and Engineering, The Pennsylvania State University, University Park, Pennsylvania 16802, USA}
\author{N. Samarth}
\email{nsamarth@psu.edu}
\affiliation{Materials Research Institute, The Pennsylvania State University, University Park, Pennsylvania 16802, USA}\affiliation{Department of Physics, The Pennsylvania State University, University Park, Pennsylvania 16802, USA}

\date{\today}

\begin{abstract}
We report four probe measurements of the low field magnetoresistance in single core/shell GaAs/MnAs nanowires synthesized by molecular beam epitaxy, demonstrating clear signatures of anisotropic magnetoresistance that track the field-dependent magnetization. A comparison with micromagnetic simulations reveals that the principal characteristics of the magnetoresistance data can be unambiguously attributed to the nanowire segments with a zinc blende GaAs core. The direct correlation between magnetoresistance, magnetization and crystal structure provides a powerful means of characterizing individual hybrid ferromagnet/semiconductor nanostructures.

\end{abstract}

\maketitle

The incorporation of spin-related functionality into semiconductor nanostructures provides an exciting new route for nanospintronic devices. \cite{Awschalom:2007wl} Nanodevices derived from MnAs/GaAs heterostructures present an interesting opportunity in this context because GaAs is an important semiconductor for optoelectronics, while MnAs is a ferromagnetic metal with a Curie temperature above room temperature ($\sim$313-350 K, depending on the strain). Indeed, MnAs/GaAs heterostructures have excellent compatibility with commonly used semiconductor devices. \cite{Tanaka:SST02,Daweritz:RPP06} In addition, MnAs is a fundamentally interesting ferromagnet because of the unique competing interplay between the magnetocrystalline anisotropy and the shape anisotropy. \cite{EngelHerbert:PRB07} Recent work has shown that the heteroepitaxy of MnAs on GaAs can also be realized in core/shell nanowires (NWs).\cite{hilse:APL09,Dellas:APL10,Takagaki:IOP11} Such NWs constitute a novel arena for studying magnetization dynamics in restricted nanoscale geometries. However, probing the magnetization in such individual NWs is a challenge for conventional magnetometry techniques. Here, we use magnetoresistance (MR) measurements of single NW devices in conjunction with micromagnetic simulations to gain insights into the magnetization switching process of core/shell GaAs/MnAs NWs. The methodology presented here is also applicable to other hybrid core/shell semiconductor/ferromagnet NWs of current interest.\cite{Rudolph:2009fk,Gao:2011uq,Butschkow:arxiv2011}

The core/shell NW samples studied here were synthesized on GaAs (111)B substrates in an EPI 930 molecular beam epitaxy chamber. We used a catalyst-free growth technique for the GaAs NWs, \cite{Colombo:PRB08} followed by thin film growth of a MnAs shell, as detailed in an earlier report.\cite{Dellas:APL10} This growth technique contrasts with other approaches wherein GaAs/MnAs core/shell NWs are synthesized using a Au catalyst. \cite{hilse:APL09,Takagaki:IOP11} Additionally, we note that the epitaxial orientation relationship between GaAs and MnAs is different from NWs synthesized using a Au catalyst due to the different crystal structures of the GaAs core. This creates a difference in magnetocrystalline anisotropy that has a significant impact on the magnetic domain structure of the MnAs shell and the low field magnetotransport properties.

Figure 1(a) shows a cross-sectional transmission electron microscope image of a single core/shell NW with a zinc-blende (ZB) GaAs core of $\sim$200 nm diameter. The MnAs shell thickness is estimated to be $\sim$10 nm. The GaAs core is mostly in the ZB structure with small segments of the wurtzite (WZ) phase; the MnAs shell is crystalline with a hexagonal NiAs structure. For the segments of the NW with ZB core the growth direction is along the $[111]$ direction with six facets belonging to the $\{110\}$ family. The c-axis (hard axis) of MnAs lies in plane with the NW facets, at an angle of $\sim$$\pm 53^{\circ}$ with respect to the wire axis. The c-axis of MnAs mirrors itself on adjacent facets. For the WZ part of the NW, the growth direction is along [001] and the c-axis of MnAs is along the NW axis.\cite{Dellas:APL10}

\begin{figure}[]
\includegraphics[width=3.5in]{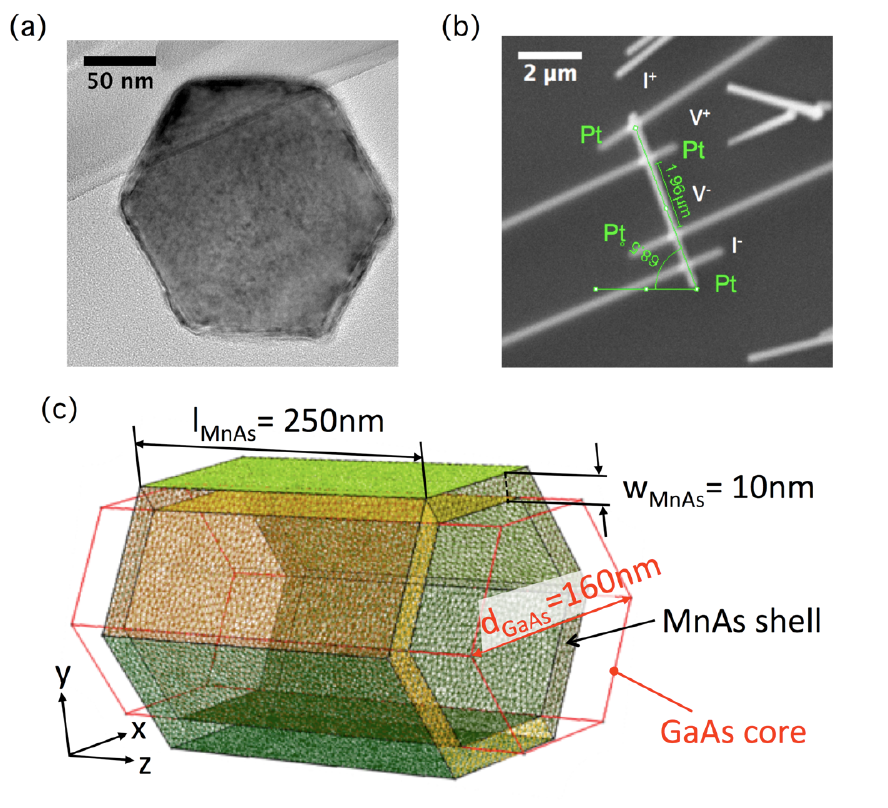}
\caption{(a) Cross-sectional TEM image of a GaAs/MnAs core/shell NW. The GaAs core is in the ZB phase. (b) Single GaAs/MnAs core/shell NW device contacted by four FIB-assisted Pt electrodes, with L = 1.96 $\mu$m. (c) GaAs/MnAs core/shell structure used in micromagnetic simulations.}
\label{fig:1}
\end{figure}

We ultrasonically removed the GaAs/MnAs core/shell NWs from the substrate and dispersed them onto a Si/Si$_3$N$_4$ substrate. The sample was then transferred into a dual-beam focused ion beam (FIB) system (FEI Quanta 200 3D) with {\it in situ} scanning electron microscopy (SEM) capabilities. After oxide layer milling, we deposited four Pt electrodes on single GaAs/MnAs core/shell NWs for electrical measurements. We minimized the Ga$^{+}$ ion imaging time to reduce contamination and also kept the Ga$^{+}$ ion deposition current and chamber pressure low to minimize the spreading of Pt. Figure 1(b) shows an SEM image of a typical device. Subsequent electrical transport measurements were carried out in a Quantum Design Physical Properties Measurement system using a standard four-probe AC resistance bridge. We made sure that the contacts are ohmic and used a typical excitation current of 0.5 $\mu$A. We measured the MR over a temperature range $500$~mK to $300$~K and in magnetic fields up to $80$~kOe. In total, we fabricated and measured four devices. In this Letter, we focus on data from only one of these devices; the other devices show qualitatively similar behavior. In addition, we carried out control measurements using a bare GaAs NW device (i.e. without any MnAs shell) using the same FIB contacting technique with Pt electrodes. This control experiment shows that the nominally undoped GaAs core is highly insulating. Thus, the GaAs core merely serves as a NW template that supports the metallic MnAs shell which dominates the transport data discussed in this paper.

\begin{figure}[]
\includegraphics[width=3.5in]{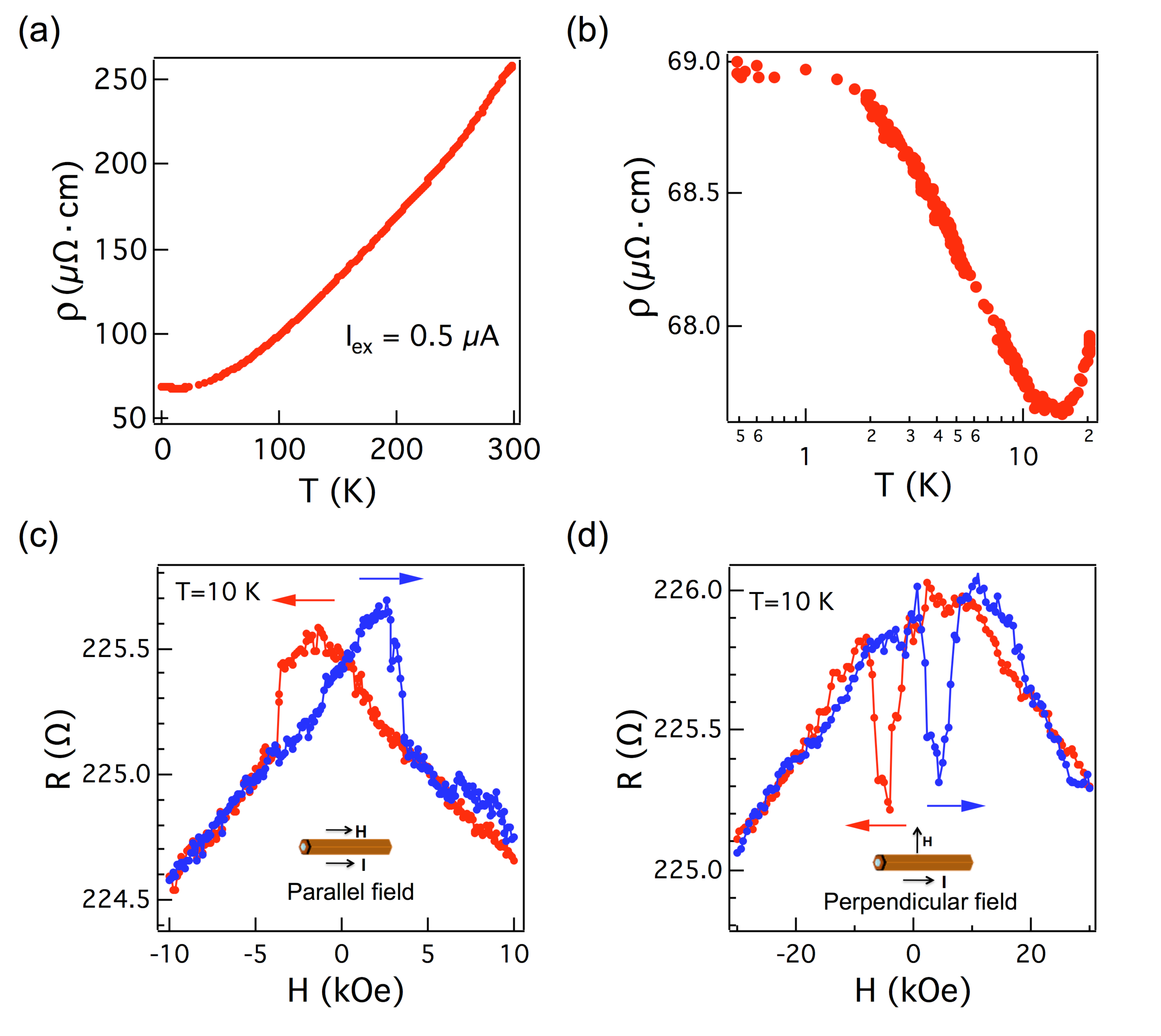}
\caption{(a) Resistivity $\rho$ of the device in Fig. 1(b) as a function of the temperature with an excitation current of 0.5 $\mu$A. (b) Resistivity vs. temperature (plotted on a log scale). Experimental MR loops with a magnetic field applied (c) parallel and (d) perpendicular to wire axis.}
\label{fig:2}
\end{figure}

Figure 2(a) shows the temperature-dependent resistivity of a GaAs/MnAs core/shell NW with length $\sim$1.96 $\mu$m from room temperature ($\sim$298 K) down to $\sim$500 mK. The resistivity of the MnAs shell was $\sim 5 \times 10^{-4} ~\Omega\cdot$cm at room temperature, which is $\sim$5 times higher than a typical MnAs epilayer grown on GaAs(001). The temperature dependence of the resistivity shows metallic behavior, similar to that of a MnAs epilayer between room temperature and 20 K. However, the somewhat smaller residual resistivity ratio (defined as the ratio of the resistivity at 300 K to that at 4.2 K) indicates that these MnAs shells are more disordered than epitaxial films of similar thicknesses. \cite{Berry:PRB01} Below $T\approx15~K$, the resistivity increases with decreasing temperature (Fig. 2(b)). Analysis to be reported elsewhere shows a temperature dependence of the {\it conductivity} $\sigma \sim \ln T$, consistent with the onset of localization in a diffusive two dimensional system. The saturation of the resistivity at even lower temperatures ($T \lesssim 1.4$~K), shown in Fig.\ 2(b), is not yet understood, but could arise either from trivial heating effects or from more interesting dimensional crossover as the relevant length scales (such as the phase breaking length) increase with lowering temperature.

Figures 2(c) and 2(d) show the MR of a NW device measured at low temperature ($T = 10$ K) in magnetic fields applied parallel and perpendicular to the wire axis, respectively. These measurements were carried out after first saturating the magnetization of the MnAs shell at 80 kOe. We observe a hysteretic MR at low fields, superimposed upon a linear negative MR that does not saturate even at the highest fields used in the present measurements (high field data not shown). The physical origin of this interesting non-saturating MR will be discussed elsewhere. We focus here on the low field MR. For magnetic fields along the wire axis, as we decrease the magnitude of the field from its maximum value to zero, the resistance initially increases (linear background) and reaches a maximum after field reversal at $\sim$3.5 kOe (Fig.\ 2(c)). The MR then abruptly changes, indicating domain switching. From the epitaxial relationship between the MnAs shell and the GaAs core, we conclude that this MR feature originates from wire segments with a ZB core. Our reasoning is as follows: for segments with a WZ core, the MnAs hard axis is along the wire axis; the strong magnetocrystalline anisotropy energetically disfavors magnetization in that direction and changing the applied field in that direction would not give rise to abrupt changes in the MR. Figure 2(d) shows the MR with the magnetic field perpendicular to the wire axis. Here, the MR is more complex: as we decrease the magnitude of the field from its maximum value to zero, the resistance again initially increases (linear background); the resistance then drops sharply after field reversal to a minimum at {$\sim$8} kOe before recovering at {$\sim$4} kOe. This behavior suggests a two step reversal process. Again, contributions from wire segments with a WZ core cannot cause the abrupt changes observed in the MR data: the magnetization would be exclusively in the cross section of the WZ wire segments favored by both the external applied field and magnetocrystalline anisotropy. 

Both hysteretic effects in the measured MR loops persist up to room temperature and are classic signatures of anisotropic magnetoresistance (AMR). Since AMR is directly connected to the magnetization of a ferromagnetic sample, we can exploit the effect as a sensitive probe of the field induced rotation and switching of the magnetization in these nanostructures.\cite{Wegrowe:PRL99} To further gain insight into the magnetization reversal process of GaAs/MnAs NWs we carried out micromagnetic simulations. The magnetic domain structure of MnAs thin films has already been the subject of micromagnetic studies using finite difference based solvers.\cite{EngelHerbert:PRB07, EngelHerbert:APL06} This approach however is not suitable due to the geometry of the core/shell structure. We therefore employed the open source finite element code MAGPAR \cite{Scholz:CMS03} to avoid erroneous results from the staircase approximation.\cite{Donahue:IEEE07} The domain configuration was calculated using the Landau-Lifshitz-Gilbert equation with the damping constant $\alpha = 0.1$. The following micromagnetic parameters for MnAs were used: exchange stiffness constant $A = 1\times 10^{-11}$~J/m, uniaxial magnetocrystalline anisotropy constant $K_{u} = - 7.2 \times 10^{5}$~J/m$^{3}$, and saturation magnetization $M_{s} = 8 \times 10^{5}$~A/m. These parameters were successfully employed previously to simulate the magnetic domain structure of MnAs thin films grown on GaAs(001) \cite{EngelHerbert:PRB07} and on GaAs(111) substrates. \cite{EngelHerbert:APL06} We varied MnAs shell thickness and GaAs core diameter from $10$ to $20$~nm and from $120$ to $160$~nm, respectively. Here we present the results of a $10$~nm thick MnAs shell on a ZB GaAs core ($160$~nm in diameter) being closest to the  NW geometry investigated. The core/shell NW geometry caused a very large boundary element matrix and we therefore limited the wire length to $250$~nm. The finite element mesh was generated using GMESH.\cite{NME:NME2579} The average edge length of the tetrahedral elements around $5.5$~nm was chosen close to the micromagnetic exchange lengths and the field stepping was as small as $25$~Oe.

Figure 3(a) shows the simulated hysteresis curve with the magnetic field applied along the wire axis. The hysteresis shows a single domain reversal with a coercive field of $4.85$~kOe in good agreement with the measurements. The energy barrier separating the two stable domain states is caused by the magnetocrystalline hard axis being inclined with the wire axis. We used the simulation results as the input to the standard heuristic description of AMR: \cite{McGuire:MagIEEE}

\begin{equation}
\label{eqnAMR}
\rho=\rho_{\perp}+\left( \rho_{\parallel}-\rho_{\perp} \right) \cos^2 \varphi,
\end{equation}

where $\rho_{\parallel}$ and $\rho_{\perp}$ are the longitudinal and transversal resistivities with respect to magnetization and $\varphi$ the angle between the magnetization and current density. For the NW, the angle is given by the normalized magnetization along the wire axis: $\cos\varphi=M_\parallel$. Note that for MnAs $\rho_{\parallel} < \rho_{\perp}$.\cite{Takagaki:JAP07} Figure 3(c) shows that the AMR effect calculated by Eqn.\ \ref{eqnAMR} is in very good agreement with the measured MR loop in Fig.\ 2(c), reproducing shape and coercive field well. Note that although there is a reduction in MR with increasing field, it does not explain the linear background at higher magnetic fields. The overestimation in coercive field is attributed to an underestimation of the MnAs shell width. Simulations with varying NW geometries showed a decrease in coercive field with increasing shell thickness. The experimental determination of the MnAs thickness from the TEM image was challenging due to the low contrast between the core and the shell and is held responsible for this discrepancy.

\begin{figure}[t]
\includegraphics[width=3.5in]{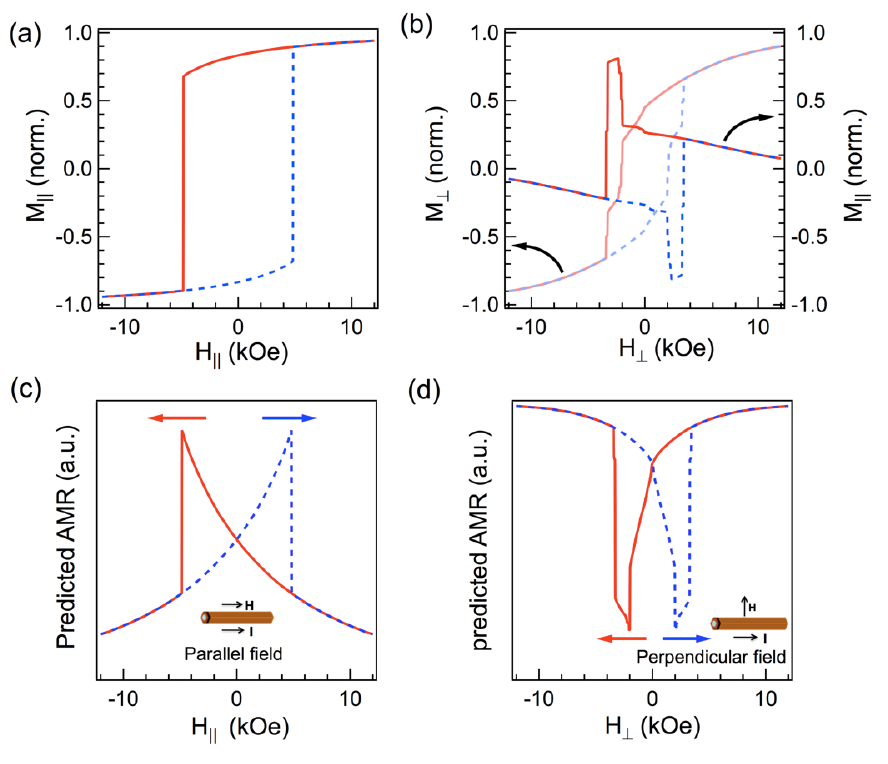}
\caption{Simulated magnetic hysteresis curves of GaAs/MnAs core shell nanowires with a ZB core and magnetic fields applied (a) along the wire axis ($z$) and (b) perpendicular to the wire axis ($x$). In panel (b), we show hysteresis loops for the magnetization component along the applied field direction $x$ and perpendicular to the wire axis $M_{\perp}$, and along the wire axis $M_{\parallel}$. AMR effect extracted from the simulated hysteresis curves for magnetic fields (c) parallel and (d) perpendicular to the wire axis.}
\label{fig:3}
\end{figure}

Figures 3(b) and 3(d) show the simulated hysteresis curves and AMR loops for magnetic fields applied perpendicular to the wire axis [$x$-axis, see Fig.\ 1(c)]. Two magnetization curves, $M_{\perp}$ in the applied field direction and $M_{\parallel}$ along the wire axis, are overlayed. The hysteresis reveals two discontinuous changes at $2.1$~kOe and $3.4$~kOe: the two facets aligned within the applied field [upper and lower facet, cf.\ Fig.\ 1(c)] reverse at a lower field, whereas all other facets that are inclined to the applied field direction switch at the higher field value at once. The increase in magnetization along the wire axis $M_{\parallel}$ between the two switching events originates from the epitaxial relationship, i.e., the inclination of the hard axis with respect to the wire axis and the alternation of the angle in adjacent facets. Reducing the magnetic field from saturation, the magnetization in the facets deviates from the $x$-direction and aligns in the respective facet planes, perpendicular to the ``local'' hard axis, to minimize demagnetization and magnetocrystalline anisotropy energies. This causes $M_{\parallel}$ of the upper and lower facet to be antiparallel with $M_{\parallel}$ of all other facets, which in turn gives rise to a small net $M_{\parallel}$ of the wire. Reversing $M_{\perp}$ is accompanied with reversing $M_{\parallel}$. If the upper and lower facets reverse, $M_{\parallel}$ has the same orientation in all facets between the two switching fields, giving rise to a large $M_{\parallel}$. At larger fields the magnetization of the slanted facets reverses, reducing the net $M_{\parallel}$. The calculated switching fields are smaller than measured. A possible source of error is the alignment of the NW in the applied field. Although a sufficiently precise line-up of the nanowire axis was easily achieved, the control over the azimuthal orientation is challenging and might cause the discrepancy. 
Disregarding the linear background, the measured MR is well reproduced by the AMR effect derived from the magnetization curve $M_{\parallel}$. A two-step reversal process and the reduction in MR between the two reversal field steps due to a large $M_{\parallel}$ are correctly predicted.

In conclusion, low field MR measurements of single GaAs/MnAs core/shell NWs reveal the AMR effect of the wire segments with a ZB GaAs core superimposed on a linear background. Both MR loops measured for fields perpendicular and parallel to the wire axis are well reproduced by micromagnetic simulations, even for a relatively complex geometry. The combination of MR measurements with micromagnetic simulations thus provides a powerful means to gain insight into the domain structure and dynamic properties of functional ferromagnetic nanostructures.

This work is supported by the Penn State Center for Nanoscale Science under NSF DMR-0820404, and by NSF ECCS-0609282, ONR N00014-09-1-0221, and the Penn State Nanofabrication Facility  NSF NNUN ECCS-35765.

%

\end{document}